\documentclass[graybox]{svmult}

\usepackage{type1cm}        
\usepackage{makeidx}         
\usepackage{graphicx}        
\usepackage{multicol}        
\usepackage[bottom]{footmisc}

\usepackage{newtxtext}       %
\usepackage{newtxmath}       

\makeindex             

\usepackage{amsmath}
\usepackage{amssymb} 
\usepackage{dsfont} 
\usepackage{mathrsfs} 
\usepackage{braket} 

\newcommand{\ketbra}[2]{\ensuremath{|#1 \vphantom{#2}\rangle \langle #2\vphantom{#1}|}}
\newcommand{\tr}[1]{\mathrm{Tr}\left(#1\right)}
\newcommand{\tra}[1]{\mathrm{Tr}(#1)}
\newcommand{\abs}[1]{\left| #1 \right|}

\usepackage{doi} 

\begin{document}

\title*{Complementarity between one- and two-body visibilities}

\author{Christoph Dittel and Gregor Weihs}

\institute{Christoph Dittel \at Physikalisches Institut, Albert-Ludwigs-Universit{\"a}t Freiburg, Hermann-Herder-Str. 3, 79104 Freiburg, Germany, \email{christoph.dittel@physik.uni-freiburg.de}
\and Gregor Weihs \at Institut f{\"u}r Experimentalphysik, Universit{\"a}t Innsbruck, Technikerstr. 25, 6020 Innsbruck, Austria, \email{gregor.weihs@uibk.ac.at}}

\maketitle

\abstract*{We study one- and two-body visibility measures under an optimization of common, i.e. global evolutions of a two-body system, and identify two different visibilities of two-body correlators, both behaving complementary to the usual one-body interference visibility. We show that only one of them satisfies the common inequality associated with a complementary relation, while the other one entails a contrary relation. This, however, can be understood in terms of entanglement between the constituents.}

\abstract{We study one- and two-body visibility measures under an optimization of common, i.e. global evolutions of a two-body system, and identify two different visibilities of two-body correlators, both behaving complementary to the usual one-body interference visibility. We show that only one of them satisfies the common inequality associated with a complementary relation, while the other one entails a contrary relation. This, however, can be understood in terms of entanglement between the constituents.}

\section{Introduction}

Complementarity was initially discussed \cite{Bohr-QM-1935,Bohr-DE-1949} and first made quantitative at the single particle level \cite{Greenberger-SW-1988,Jaeger-TI-1995,Englert-FV-1996}. For multiple particles the issue is obviously much more complicated, because there are a myriad possibilities of defining measures of different types of information that might be complementary or not. Yet, the single-particle case can be lifted to the two-particle level with complementarity between one- and two-particle visibility for the scenario where the two particles are subject to separate two-mode interferometers \cite{Jaeger-CO-1993,Schlienz-DE-1995}. Clearly Mike Horne wanted to expand on this and one of us (GW) distinctly remembers his talk at the conference ``Epistemological and Experimental Perspectives on Quantum Physics'' (Vienna, Austria, 1998) \cite{Horne-CF-1999} where he tried to extend complementarity to the interference of three particles. The talk showed that there is no such straightforward generalization. In a recent result \cite{Dittel-WP-2018}, we were able to derive complementarity and duality relations for multiple, partially distinguishable interfering particles but not in the way envisaged by Mike Horne. Both approaches reveal the crucial role of entanglement, and for tripartite systems it is known that entanglement measures cannot both be faithful and monogamous \cite{Lancien-EM-2016}. On the other hand, the situation is rather simple for two two-level systems (qubits) since all measures of entanglement are similar in this case. Under this perspective, we here elucidate a possible connection between the two scenarios at the two-particle level with special emphasis on the role of entanglement.

\begin{figure}[t]
\centering
\includegraphics[width=\linewidth]{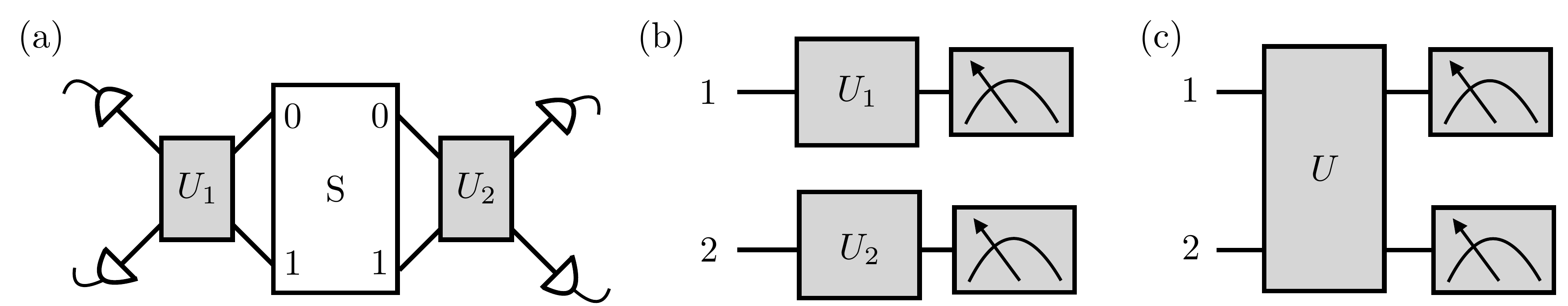}
\caption{Schematic experimental settings to probe the complementarity of one- and two-body visibilities. (a) A source $\mathrm{S}$ emits two particles, which are possibly entangled in their mode occupation, with modes labelled $0$ and $1$. The particles leave the source in opposite directions, are separately transformed according to $U_1$ and $U_2$, and measured in the output modes, respectively. (b) In a similar setting as in (a), two possibly entangled qubits, labelled $1$ and $2$, undergo local unitary transformations with a subsequent measurement in the computational basis. (b) In a similar setting as in (b), two qubits are transformed according to a common (global) two-qubit unitary transformation $U$.}
\label{fig:Experiment}
\end{figure}

The original scheme involves a two-particle source, with the particles sent in opposite directions, and possibly entangled in their occupation in two distinct modes [see Fig.~\ref{fig:Experiment}(a)]. The particles are then separately transformed according to two-mode unitary transformations [$U_1$ and $U_2$ in Fig.~\ref{fig:Experiment}(a)], and subsequently measured in the output modes. Thereby, one- and two-body visibilities are obtained after an optimization procedure over the local unitary transformations $U_1$ and $U_2$, which were shown to be in a complementary relation \cite{Jaeger-CO-1993,Jaeger-TI-1995}.

By considering the mode occupation of each particle as a two-level system, this experimental scheme is similar to two possibly entangled qubits undergoing local unitary transformations followed by a measurement in the computational basis [compare Figs.~\ref{fig:Experiment}(a) and~(b)]. Now, by allowing for common  -- i.e. global -- two-qubit unitaries instead [see Fig.~\ref{fig:Experiment}(c)], we can ask whether similar complementarity relations between one- and two-body visibilities can be obtained.

\section{One-body visibility}
To set the stage, let us consider a pure state $\ket{\Psi}$ of two qubits, which we write in it's Schmidt decomposition, $\ket{\Psi}=\sum_{j=0}^1 \sqrt{\lambda_j} \ket{\eta_j}\ket{\xi_j}$, with $\lambda_j\in[0,1]$, $\lambda_0 + \lambda_1=1$, and $\{ \ket{\eta_j} \}_{j=0}^1$ and $\{ \ket{\xi_j} \}_{j=0}^1$ an orthonormal basis of the first and second qubit, respectively. Note that w.l.o.g. we assume $\lambda_0 \geq \lambda_1$, such that $1/2 \leq \lambda_0 \leq 1$. The reduced state of the first (resp. second) qubit is then obtained from the two-qubit density operator $\rho=\ketbra{\Psi}{\Psi}$ by tracing over the Hilbert space of the second (resp. first) qubit, $\rho_1=\sum_{j=0}^1\lambda_j \ketbra{\eta_j}{\eta_j}$ (resp. $\rho_2=\sum_{j=0}^1\lambda_j \ketbra{\xi_j}{\xi_j}$), where $\lambda_0$ and $\lambda_1$ appear as the eigenvalues of $\rho_1$ (resp. $\rho_2$).

First, let us consider the reduced state $\rho_1$ of the first qubit under local unitary transformations $U_1$ as shown in Fig.~\ref{fig:Experiment}(b). In this setting, the one-body visibility is commonly written as
\begin{align}\label{eq:v1}
v_1=\frac{p_1^\mathrm{max}(0)-p_1^\mathrm{min}(0)}{p_1^\mathrm{max}(0)+p_1^\mathrm{min}(0)},
\end{align}
with $p_1^\mathrm{max}(0)=\max_{U_1} \tra{\ketbra{0}{0} U_1 \rho_1 U_1^\dagger}$ (resp. $p_1^\mathrm{min}(0)=\min_{U_1} \tra{\ketbra{0}{0} U_1 \rho_1 U_1^\dagger}$) the maximal (resp. minimal) probability to find the first qubit in $\ket{0}$ after an optimization over $U_1$. Now, for $U_1^\mathrm{max}$ the unitary leading to $p_1^\mathrm{max}(0)$, we see that the unitary $\sigma_x U_1^\mathrm{max}$, with the Pauli matrix $\sigma_x$ flipping $\ket{0}$ and $\ket{1}$,  gives rise to $p_1^\mathrm{max}(1)=p_1^\mathrm{max}(0)$. Moreover, with $p_1^\mathrm{min}(0)=1-p_1^\mathrm{max}(1)$, we have $p_1^\mathrm{min}(0)=1-p_1^\mathrm{max}(0)$, such that Eq.~\eqref{eq:v1} becomes
\begin{align}\label{eq:v1p1max}
v_1=2 p_1^\mathrm{max}(0)-1.
\end{align}
Under this perspective, we may also express the one-body visibility~\eqref{eq:v1} in terms of the \textit{Kolmogorov distance} (or $L_1$ \textit{distance}) \cite{Nielsen-QC-2011}
\begin{align}\label{eq:KolmoDist}
D(P_1,P_1^\mathrm{mix})=\frac{1}{2} \sum_{j=0}^1\abs{p_1(j)-\frac{1}{2}}
\end{align}
between the probability distributions $P_1=\{p_1(0),p_1(1)\}$ and $P^\mathrm{mix}_1=\{1/2,1/2\}$. The latter is obtained from a maximally mixed one-qubit state $\rho_1^\mathrm{mix}=1/2 \sum_{j=0}^1 \ketbra{\eta_j}{\eta_j}$. In particular, with the help of Eqs.~\eqref{eq:v1p1max} and~\eqref{eq:KolmoDist}, we find
\begin{align}\label{eq:v1D}
v_1&=\max_{U_1} 2\ D(P_1,P_1^\mathrm{mix})\\
&=\max_{U_1}\sum_{j=0}^1 \abs{\tr{\ketbra{j}{j} U_1 (\rho_1-\rho_1^\mathrm{mix}) U_1^\dagger} }.
\end{align}
Here, the maximum is reached if $U_1^\dagger\ketbra{0}{0}U_1$ and $U_1^\dagger\ketbra{1}{1}U_1$ project onto the  eigenstates of $\rho_1-\rho_1^\mathrm{mix}$, e.g. $U_1^\dagger\ket{0}=\ket{\eta_0}$ and $U_1^\dagger\ket{1}=\ket{\eta_1}$, such that $p_1^\mathrm{max}(0)=\lambda_0$. Together with Eq.~\eqref{eq:v1p1max}, we then arrive at
\begin{align}\label{eq:v1Eigenvalue}
v_1=2 \lambda_0-1.
\end{align}
This expression allows us to write $v_1$ as the normalized purity \cite{Dittel-WP-2018} of the reduced one-qubit state,
\begin{align}\label{eq:v1tr}
v_1=\sqrt{2\tr{\rho_1^2}-1},
\end{align}
or in terms of the concurrence $\mathcal{C}=2\sqrt{\lambda_0\lambda_1}$ \cite{Wootters-EF-1998} of $\rho$, which quantifies the entanglement between the qubits,
\begin{align}\label{eq:v1C}
v_1 = \sqrt{ 1-\mathcal{C}^2 }.
\end{align}
Accordingly, maximally entangled qubits, for which $\mathcal{C}=1$, imply a vanishing visibility $v_1=0$, and separable qubits, for which $\mathcal{C}=0$, lead to $v_1=1$. Given that the concurrence $\mathcal{C}$ is an entanglement measure, it is apparent from Eq.~\eqref{eq:v1C} that the one-body visibility $v_1$ quantifies the separability of the qubits, and, thus, stands in a complementary relation to the entanglement between the qubits. In this regard, we now focus on visibility measures of two-body correlators that quantify the amount of entanglement under an optimization of global two-qubit unitary transformations $U$ as illustrated in Fig.~\ref{fig:Experiment}(c).

\section{Two-body visibilities}

Previous works \cite{Jaeger-CO-1993,Jaeger-TI-1995,Schlienz-DE-1995,Horne-CF-1999} considered the experimental schemata of Fig.~\ref{fig:Experiment}(a) or~\ref{fig:Experiment}(b), with the two-body visibility
\begin{align}\label{eq:v12}
v_{12}=\frac{\bar{p}^\mathrm{loc.max}(0,0)-\bar{p}^\mathrm{loc.min}(0,0)   }{ \bar{p}^\mathrm{loc.max}(0,0)+\bar{p}^\mathrm{loc.min}(0,0)  }
\end{align}
obtained under optimization of the two-body correlator
\begin{align}\label{eq:pbar}
\bar{p}(j,k)=p(j,k)-p_1(j)p_2(k)+1/4
\end{align}
with respect to local unitary transformations $U_1$ and $U_2$ [indicated by the superscript $\mathrm{loc.max}$ and $\mathrm{loc.min}$]. In this case, $p(j,k)=\tra{\ketbra{j,k}{j,k} (U_1\otimes U_2) \rho (U_1\otimes U_2)^\dagger}$ is the probability to measure the two-qubit state $ \ket{j,k} \equiv \ket{j}\ket{k}$, and the constant $1/4$ is needed in order that $\bar{p}(j,k) \geq 0$ and $\sum_{j,k=0}^1 \bar{p}(j,k)=1$, such that $\bar{P}=\{\bar{p}(j,k)\}_{j,k=0}^1$ has the properties of a probability distribution. As shown in Ref.~\cite{Jaeger-TI-1995}, Eq.~\eqref{eq:v12} results in $v_{12}=\mathcal{C}$, such that, by Eq.~\eqref{eq:v1C}, we have
\begin{align}\label{eq:v1v12=}
v_1^2+v_{12}^2=1.
\end{align}
Let us note that in Ref.~\cite{Jaeger-CO-1993} the unitaries $U_1$ and $U_2$ are restricted to transformations corresponding to a balanced beam splitter, having equal transmittivity and reflectivity, together with a phase shifter in one of the input modes. Under these restrictions, Eq.~\eqref{eq:v1v12=} becomes \cite{Jaeger-CO-1993,Jaeger-TI-1995}
\begin{align}\label{eq:v1v12}
v_1^2+v_{12}^2\leq 1.
\end{align}

By performing the maximization in Eq.~\eqref{eq:v12} over \emph{global} two-body unitary transformations $U$ instead of local unitaries of the form $U_1 \otimes U_2$ [see Fig.~\ref{fig:Experiment}(c)], one can show that $v_{12}=1$, independently on the initial two-body state $\rho$. That is, $v_{12}$ does not provide any information about $\rho$ once we allow for global two-body unitaries $U$. Nonetheless, by modifying the correlators~\eqref{eq:pbar} as
\begin{align}\label{eq:cbar}
\bar{c}(j,k)=p(j,k)-p^\mathrm{sep}(j,k)+1/4,
\end{align}
we can construct entanglement sensitive two-body visibility measures of the form~\eqref{eq:v12}, with $\bar{c}(j,k)$ coinciding with $\bar{p}(j,k)$ from Eq.~\eqref{eq:pbar} in the case of local unitary transformations. While here, the usual probability $p(j,k)=\tra{\ketbra{j,k}{j,k} U \rho U^\dagger}$ is obtained from the two-body state $\rho$, which possibly involves entanglement between the qubits, $p^\mathrm{sep}(j,k)=\tra{\ketbra{j,k}{j,k} \allowbreak U \rho^\mathrm{sep} U^\dagger}$ results from the separable, and, thus, unentangled state $\rho^\mathrm{sep}=\rho_1\otimes \rho_2$. From this perspective, the correlators $\bar{c}(j,k)$ are sensitive to the state's separability, and, as will be shown further down, can be utilized in order to measure the entanglement between the qubits. Note that again the factor $1/4$ ensures $\bar{c}(j,k)\geq 0$, and the set $\bar{C}=\{\bar{c}(j,k)\}_{j,k=0}^1$ has the properties of a probability distribution. Moreover, let us recall that $\rho_1$ (resp. $\rho_2$) is obtained from $\rho$ by tracing out the second (resp. first) qubit. In the laboratory, the state $\rho^\mathrm{sep}=\rho_1\otimes \rho_2$ can then be prepared, for example, by assembling the two-qubit system from two separate copies of $\rho$, with one qubit taken from each copy.

\begin{figure}[t]
\centering
\includegraphics[width=\linewidth]{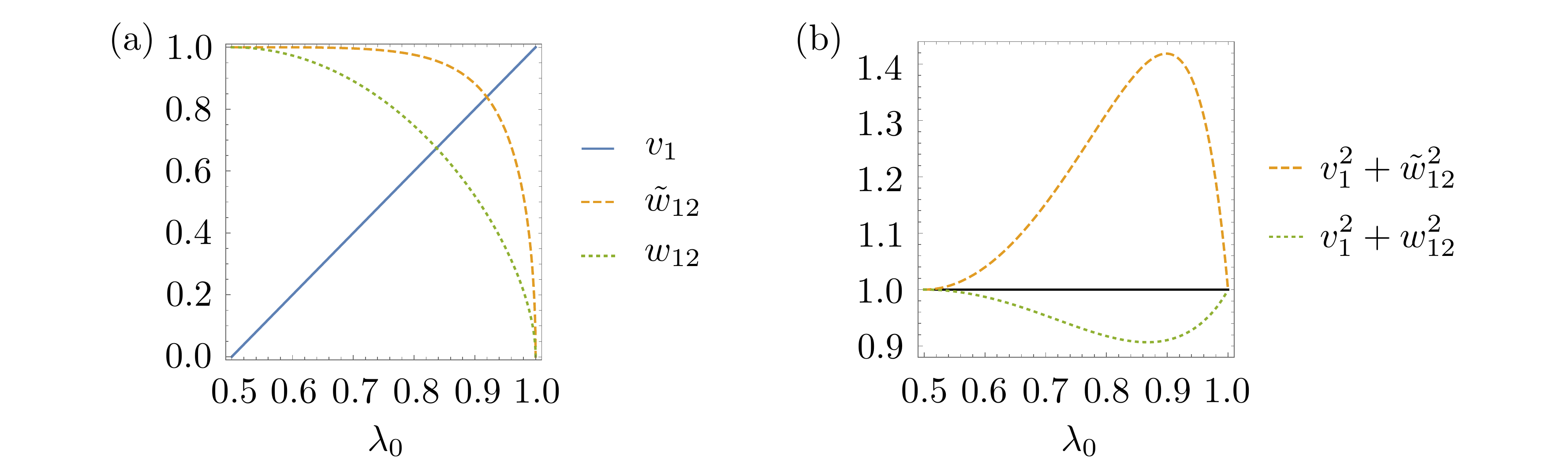}
\caption{Complementary behaviour of the visibility measures. (a) The one-body visibility $v_1$ (full blue line) monotonously increases, and the two-body visibilities $\tilde{w}_{12}$ (dashed orange line) and $w_{12}$ (dotted green line) monotonously decrease as a function of the eigenvalue $\lambda_0$. This illustrates the complementary behaviour between $v_1$ and $\tilde{w}_{12}$ (resp. $w_{12}$). (b) The sum of squared visibility measures $v_1^2+\tilde{w}_{12}^2$ (resp. $v_1^2+w_{12}^2$) is shown by an dashed orange (resp. dotted green) line as a function of $\lambda_0$, and, in accordance with inequality~\eqref{eq:v1w12tilde} (resp.~\eqref{eq:v1w12}), bounded from below (resp. above) by unity (full black line).}
\label{fig:Graph-Proof}
\end{figure}

Let us first start out from a similar definition for the visibility as in Eq.~\eqref{eq:v12}, and consider
\begin{align}\label{eq:w12tilde}
\tilde{w}_{12}= \frac{\bar{c}^\mathrm{max}(0,0)-\bar{c}^\mathrm{min}(0,0)   }{ \bar{c}^\mathrm{max}(0,0)+\bar{c}^\mathrm{min}(0,0)  },
\end{align}
with the optimization performed over global two-qubit unitaries $U$ [cf. Fig.~\ref{fig:Experiment}(c)]. In order to obtain an expression for $\tilde{w}_{12}$ in terms of the eigenvalue $\lambda_0$ (i.e. similar to Eq.~\eqref{eq:v1Eigenvalue}), we consider
\begin{align}
\bar{c}^\mathrm{max}(0,0)&=\max_{U} [p(0,0)-p^\mathrm{sep}(0,0)+1/4] \nonumber \\
&=\max_{U} \left[\tr{ \ketbra{00}{00} U (\rho-\rho_1\otimes\rho_2) U^\dagger}\right]+1/4. \label{eq:pbarmax}
\end{align}
Using the spectral decomposition $\rho-\rho_1\otimes \rho_2=\sum_{j=0}^3 \alpha_j \ketbra{\alpha_j}{\alpha_j}$, with not necessarily positive eigenvalues $\alpha_j$ and their corresponding eigenvectors $\ket{\alpha_j}$, we see that the maximum in~\eqref{eq:pbarmax} is obtained if $U^\dagger \ketbra{00}{00}U$  projects onto the subspace spanned by the eigenvector(s) corresponding to the maximal eigenvalue(s) $\alpha^\mathrm{max}$, i.e. for $U^\dagger\ket{00}=\ket{\alpha^\mathrm{max}}$ we have $\bar{c}^\mathrm{max}(0,0)=\alpha^\mathrm{max}+1/4$. Similar, for $\bar{c}^\mathrm{min}(0,0)$ in Eq.~\eqref{eq:w12tilde}, the minimization gives rise to the minimal eigenvalue(s) $\alpha^\mathrm{min}$, leading to $\bar{c}^\mathrm{min}(0,0)=\alpha^\mathrm{min}+1/4$. Therefore, we simply have to evaluate the eigenvalues of $\rho-\rho_1\otimes \rho_2$, for which we find
\begin{align}\label{eq:eigenvalues}
\alpha_0=\alpha_1=-\lambda_0\lambda_1, \quad \alpha_2=\lambda_0\lambda_1+\sqrt{\lambda_0\lambda_1}, \quad\alpha_3=\lambda_0\lambda_1-\sqrt{\lambda_0\lambda_1}.
\end{align}
Under consideration of $\lambda_0\lambda_1\leq 1$, we have $\alpha^\mathrm{max}=\alpha_2$ and $\alpha^\mathrm{min}=\alpha_3$, such that Eq.~\eqref{eq:w12tilde} becomes
\begin{align}\label{eq:w12tildeLambda}
\tilde{w}_{12}= \frac{2\sqrt{\lambda_0 \lambda_1}}{2\lambda_0 \lambda_1 + 1/2}.
\end{align}
This can also be expressed in terms of the concurrence $\mathcal{C}$ of the two-body state $\rho$, reading
\begin{align}\label{eq:w12tildeC}
\tilde{w}_{12}=  \frac{2 \mathcal{C}}{\mathcal{C}^2+1}.
\end{align}
From Eqs.~\eqref{eq:w12tildeLambda} and~\eqref{eq:w12tildeC} we see that $0\leq \tilde{w}_{12} \leq 1$ measures the entanglement between the qubits, with its lower bound reached for separable qubits, i.e. $\tilde{w}_{12}=0$ for $\mathcal{C}=0$, and its upper bound reached for maximally entangled qubits, i.e. $\tilde{w}_{12} = 1$ for $\mathcal{C}=1$. Therefore, $\tilde{w}_{12}$ behaves complementary to the single-body visibility $v_1$ from Eq.~\eqref{eq:v1Eigenvalue} as highlighted by their opposite monotonicity as a function of $\lambda_0$ shown in Fig.~\ref{fig:Graph-Proof}(a). At the same time, however, the visibilities $v_1$ and $\tilde{w}_{12}$ do \emph{not} satisfy the complementary relation of the form~\eqref{eq:v1v12}, instead,
\begin{align}\label{eq:v1w12tilde}
v_1^2 + \tilde{w}_{12}^2 \geq 1.
\end{align}
This inequality results from plugging $\tilde{w}_{12} \geq \mathcal{C}$ [see Eq.~\eqref{eq:w12tildeC}] into Eq.~\eqref{eq:v1C}, and is graphically illustrated in Fig.~\ref{fig:Graph-Proof}(b). While Eq.~\eqref{eq:v1w12tilde} first seems counter intuitive for two normalized and complementary behaving measures, it highlights that satisfying the relation of the form~\eqref{eq:v1v12} is not a sufficient and not even a necessary condition for two measures behaving complementary to each other. Nevertheless, in the following we provide an alternative visibility measure, which we will show to satisfy the form~\eqref{eq:v1v12}.

The definitions of the visibilities $v_{12}$ and $\tilde{w}_{12}$ from Eqs~\eqref{eq:v12} and~\eqref{eq:w12tilde} are motivated by their similarity to the usual one-body visibility $v_1$ from Eq.~\eqref{eq:v1}. However, as shown in Eq.~\eqref{eq:v1D}, $v_1$ can likewise be considered as the Kolmogorov distance to the outcome obtained from a maximally mixed state. In view of the two-body correlators~\eqref{eq:cbar} after global unitary transformations, Eq.~\eqref{eq:v1D} then motivates the definition of the visibility
\begin{align}\label{eq:w12}
w_{12}= \frac{4}{3}\max_{U} D(\bar{C}, \bar{C}^\mathrm{sep}),
\end{align}
with $\bar{C}^\mathrm{sep}=\{1/4,1/4,1/4,1/4 \}$ the correlator distribution in the case of a separable state $\rho=\rho^\mathrm{sep}$. Note that the factor $4/3$ arises in order for $w_{12}$ to be normalized. This will get apparent further down. In consideration of the correlators~\eqref{eq:cbar}, $w_{12}$ from Eq.~\eqref{eq:w12} is equivalent to
\begin{align}\label{eq:w12PP}
w_{12}= \frac{4}{3}\max_{U} D(P, P^\mathrm{sep}),
\end{align}
with $P=\{p(j,k)\}_{j,k=0}^1$ and $P^\mathrm{sep}=\{p^\mathrm{sep}(j,k)\}_{j,k=0}^1$ the probability distribution obtained from $\rho$ and $\rho^\mathrm{sep}$, respectively. Let us now inspect Eq.~\eqref{eq:w12PP}. By plugging in the Kolmogorov distance $D(P, P^\mathrm{sep})=1/2 \sum_{j,k=0}^1 |p(j,k)-p^\mathrm{sep}(j,k)|$, we obtain
\begin{align}
w_{12}=\frac{2}{3} \max_{U} \sum_{j,k=0}^1 \abs{ \tr{\ketbra{j,k}{j,k} U (\rho-\rho_1\otimes \rho_2) U^\dagger }  }.\label{eq:w12Sum}
\end{align}
Here, the maximum is reached for $U^\dagger$ rotating the computational basis $\{\ket{j,k}\}_{j,k=0}^1$ to the eigenbasis $\{\ket{\alpha_j}\}_{j=0}^3$ of $\rho-\rho_1\otimes \rho_2$, e.g. if $U^\dagger\ket{00}=\ket{\alpha_0}$, $U^\dagger\ket{01}=\ket{\alpha_1}$, $U^\dagger\ket{10}=\ket{\alpha_2}$, and $U^\dagger\ket{11}=\ket{\alpha_3}$. Equation~\eqref{eq:w12Sum} then becomes $w_{12}= 2/3  \sum_{j=0}^3 \abs{\alpha_j}$, and, with the eigenvalues~\eqref{eq:eigenvalues}, we arrive at
\begin{align}\label{eq:w12Eigenvalues}
w_{12}=\frac{4}{3} \left( \lambda_0 \lambda_1 + \sqrt{\lambda_0\lambda_1}\right).
\end{align}
Let us note that we can further express $w_{12}$ in terms of the fidelity $F(\rho_1,\rho_1^\mathrm{mix})=\sum_{j=0}^1 \sqrt{\lambda_j/2}$ of $\rho_1$ and the maximally mixed one-qubit state $\rho_1^\mathrm{mix}$ [see below Eq.~\eqref{eq:KolmoDist}],
\begin{align}\label{eq:w12F}
w_{12}=\frac{4}{3} \left[ F^4(\rho_1,\rho_1^\mathrm{mix}) -\frac{1}{4} \right],
\end{align}
where $1/4 \leq F^4(\rho_1,\rho_1^\mathrm{mix}) \leq 1$, as well as in terms of the concurrence $\mathcal{C}$ of $\rho$,
\begin{align}\label{eq:w12C}
w_{12}=\frac{1}{3} \left( \mathcal{C}^2+2\mathcal{C} \right).
\end{align}
Equations~\eqref{eq:w12Eigenvalues}-\eqref{eq:w12C} show that the two-body visibility $w_{12}$ is a normalised measure of the entanglement between the qubits, with $w_{12}=0$ for separable two-qubit states and $w_{12}=1$ for maximally entangled states. As shown in Fig.~\ref{fig:Graph-Proof}(a), $w_{12}$ monotonously decreases for increasing $\lambda_0$, obeying a complementary behaviour to the single-body visibility $v_1$ from Eq.~\eqref{eq:v1}. Furthermore, by plugging $w_{12} \leq \mathcal{C}$ [see Eq.~\eqref{eq:w12C}] into Eq.~\eqref{eq:v1C}, we find these measures to satisfy
\begin{align}\label{eq:v1w12}
v_1^2 + w_{12}^2 \leq 1,
\end{align}
which is graphically illustrated in Fig.~\ref{fig:Graph-Proof}(b). There one can also see that the inequality in~\eqref{eq:v1w12} saturates only if $v_1$ or $w_{12}$ equals unity. In summary, the one- and two-body visibilities $v_1$ and $w_{12}$ satisfy the usual relation~\eqref{eq:v1w12} associated with complementary measures, which, in the present case, can simply be interpreted in terms of entanglement: the one-body visibility $v_1$ measures the separability, and the two-body visibility $w_{12}$ the entanglement between both qubits. The more separable the qubits, the less entangled they are, and vice versa.

\section{Discussion and conclusion}
In the literature, quantitative expressions for the complementarity between two measures are often provided in terms of inequalities of the form~\eqref{eq:v1v12}. One type of these celebrated inequalities connects one- and two-body visibilities obtained after local unitary transformations of two entangled two-level systems \cite{Jaeger-CO-1993,Jaeger-TI-1995,Schlienz-DE-1995,Horne-CF-1999}. Here we went one step further and considered \emph{global} two-body transformations. We introduced two different two-body visibilities $\tilde{w}_{12}$ and $w_{12}$, and showed that both behave complementary to the usual one-body visibility $v_1$ -- in the sense of an opposite monotonicity behaviour. This can ultimately be understood in terms of entanglement: $v_1$ measures the separability, and both $\tilde{w}_{12}$ and $w_{12}$ the amount of entanglement of the two-body system under consideration. However, we found that only $w_{12}$ satisfies the celebrated complementarity relation $v_1^2 +w_{12}^2 \leq 1$. For $\tilde{w}_{12}$ on the other hand, we even showed that $v_1^2 +\tilde{w}_{12}^2 \geq 1$, although both visibility measures $v_1$ and $\tilde{w}_{12}$ are normalised and complementary to each other. While this inequality appears unexpected at first sight, it shows that satisfying the complementarity relation of the form~\eqref{eq:v1v12} is not a necessary condition, and, to keep in mind, does not suffice to speak of a complementary behaviour between two measures. It is worth mentioning that the here obtained interrelations~\eqref{eq:v1w12tilde} and~\eqref{eq:v1w12} between one- and two-body visibilities are experimentally feasible with state-of-the-art technology on diverse experimental platforms, ranging from trapped ion systems to superconducting quantum circuits.

To finish, let us comment on Michael Horn's vision of a three-body complementarity relation, possibly in the form $v_1^2+v_{12}^2+v_{123}^2 \leq 1$, with $v_{123}$ a three-body visibility. This relation was disproven by himself  \cite{Horne-CF-1999} for an extension of the correlators~\eqref{eq:pbar} to three parties together with a three-body visibility measure similar to Eq.~\eqref{eq:v12}. From our above discussion, however, it is clear that such a three-body complementarity relation is preferably addressed in terms of entanglement between the constituents. A promising approach may involve visibility measures in terms of distances between correlator distributions rather than visibilities of single correlators, and possibly two- and three-body correlators similar to those from Eq.~\eqref{eq:cbar}.  Yet, a straightforward extension of our results is not possible since we started out from the Schmidt decomposition of a two-body state, which, in general, cannot be generalised to more than two parties. Nonetheless, we are confident that the here established framework provides a promising route to a quantitative study of complementarity in multi-partite systems.

\begin{acknowledgement}
C.D. would like to thank Giulio Amato and Eric Brunner for fruitful discussions. G.W. would like to thank Barbara Kraus for making us aware of the relation to entanglement monogamy. This work was supported by the Austrian Science Fund (FWF), project no. I2562.
\end{acknowledgement}
%


\begin{thebibliography}{10}
\providecommand{\url}[1]{{#1}}
\providecommand{\urlprefix}{URL }
\expandafter\ifx\csname urlstyle\endcsname\relax
  \providecommand{\doi}[1]{DOI \discretionary{}{}{}#1}\else
  \providecommand{\doi}{DOI \discretionary{}{}{}\begingroup
  \urlstyle{rm}\Url}\fi

\bibitem{Bohr-QM-1935}
N.~Bohr, Physical Review \textbf{48}, 696 (1935)

\bibitem{Bohr-DE-1949}
N.~Bohr, in \emph{The Library of Living Philosophers, Volume 7. Albert
  Einstein: Philosopher-Scientist}, ed. by P.A. Schilpp (Open Court, 1949), pp.
  199--241

\bibitem{Greenberger-SW-1988}
D.M. Greenberger, A.~Yasin, Physics Letters A \textbf{128}(8), 391  (1988)

\bibitem{Jaeger-TI-1995}
G.~Jaeger, A.~Shimony, L.~Vaidman, Physical Review A \textbf{51}, 54 (1995)

\bibitem{Englert-FV-1996}
B.G. Englert, Physical Review Letters \textbf{77}, 2154 (1996)

\bibitem{Jaeger-CO-1993}
G.~Jaeger, M.A. Horne, A.~Shimony, Physical Review A \textbf{48}, 1023 (1993)

\bibitem{Schlienz-DE-1995}
J.~Schlienz, G.~Mahler, Physical Review A \textbf{52}, 4396 (1995)

\bibitem{Horne-CF-1999}
M.~Horne, in \emph{Epistemological and Experimental Perspectives on Quantum
  Physics}, ed. by D.~Greenberger, W.L. Reiter, A.~Zeilinger (Springer
  Netherlands, Dordrecht, 1999), pp. 211--220

\bibitem{Dittel-WP-2018}
C.~Dittel, G.~Dufour, G.~Weihs, A.~Buchleitner, arXiv:1901.02810v2  (2019)

\bibitem{Lancien-EM-2016}
C.~Lancien, S.~Di~Martino, M.~Huber, M.~Piani, G.~Adesso, A.~Winter, Phys. Rev.
  Lett. \textbf{117}, 060501 (2016)

\bibitem{Nielsen-QC-2011}
M.A. Nielsen, I.L. Chuang, \emph{Quantum Computation and Quantum Information:
  10th Anniversary Edition}, 10th edn. (Cambridge University Press, New York,
  NY, USA, 2011)

\bibitem{Wootters-EF-1998}
W.K. Wootters, Physical Review Letters \textbf{80}, 2245 (1998)

\end{thebibliography}

\end{document}